\documentclass[journal=jacsat,manuscript=article]{achemso}
\SectionNumbersOn
\usepackage[version=4]{mhchem} 

\usepackage{siunitx} 
\usepackage{amsmath,amssymb,gensymb,amsfonts,latexsym,mathtools} 
\usepackage{tabularx}
\usepackage{empheq,xfrac} 
\usepackage[hidelinks,linktocpage]{hyperref} 
\usepackage{array,makecell} 
\usepackage{booktabs}	
\usepackage{multicol,multirow} 
\usepackage{dcolumn}  
\usepackage{xcolor}
\usepackage{soul}

\definecolor{markCol1}{RGB}{0,0,0} 
\definecolor{markCol2}{RGB}{0,0,0} 

\author{Jose M. Sojo-Gordillo}
\email{jose.sojo@unibas.ch}
\affiliation{University of Basel, Klingelbergstrasse 82, 4056, Basel, Switzerland}
\altaffiliation{Both authors contributed equally}

\author{Alex Rodriguez-Iglesias}
\affiliation{Institute of Microelectronics of Barcelona, Til·lers s/n, 08193, Bellaterra, Spain}
\altaffiliation{Both authors contributed equally}

\author{Dominik M. Koch}
\affiliation{University of Basel, Klingelbergstrasse 82, 4056, Basel, Switzerland}

\author{Arianna Nigro}
\affiliation{University of Basel, Klingelbergstrasse 82, 4056, Basel, Switzerland}

\author{Iñigo Martin-Fernandez}
\affiliation{Institute of Microelectronics of Barcelona, Til·lers s/n, 08193, Bellaterra, Spain}

\author{Marta Fernandez-Regulez}
\email{marta.fernandez@imb-cnm.csic.es}
\affiliation{Institute of Microelectronics of Barcelona, Til·lers s/n, 08193, Bellaterra, Spain}

\author{Marc Salleras}
\email{marc.salleras@imb-cnm.csic.es}
\affiliation{Institute of Microelectronics of Barcelona, Til·lers s/n, 08193, Bellaterra, Spain}

\author{Ilaria Zardo}
\email{ilaria.zardo@unibas.ch}
\affiliation{University of Basel, Klingelbergstrasse 82, 4056, Basel, Switzerland}

\newcommand{\thetitle}{Thermal conductivity tuning of scalable nanopatterned silicon membranes measured with a three-probe method}
\title[Manuscript]{\thetitle}
\abbreviations{The following abbreviations are used in this manuscript:
\begin{tabular}{@{}ll}
NS    & Nanostructure\\
TCR   & Temperature Coefficient of Resistance\\
MEMS  & Micro Electro-Mechanical System\\
FEA   & Finite Element Analysis\\
TEM   & Transmission Electron Microscope\\
PCB   & Printed circuit board\\
TCR   & Temperature coefficient of resistance\\
BCP   & Block copolymer\\
SOI   & Silicon on insulator\\
BOE   & Buffered oxide etch\\
RIE   & Reactive ion etching\\
\end{tabular}
}


\begin{document}

\begin{tocentry}

Phononic silicon membranes with sub-50 nm features are fabricated via mask-free block-copolymer self-assembly, enabling controlled thermal conductivity reduction. An extended three-probe technique with laser heating provides spatially resolved measurements while accounting for contact resistances. By adjusting the etch depth of the nanoholes, the thermal conductivity of suspended silicon membranes can be directly and reproducibly tuned.

\includegraphics[width=1\textwidth]{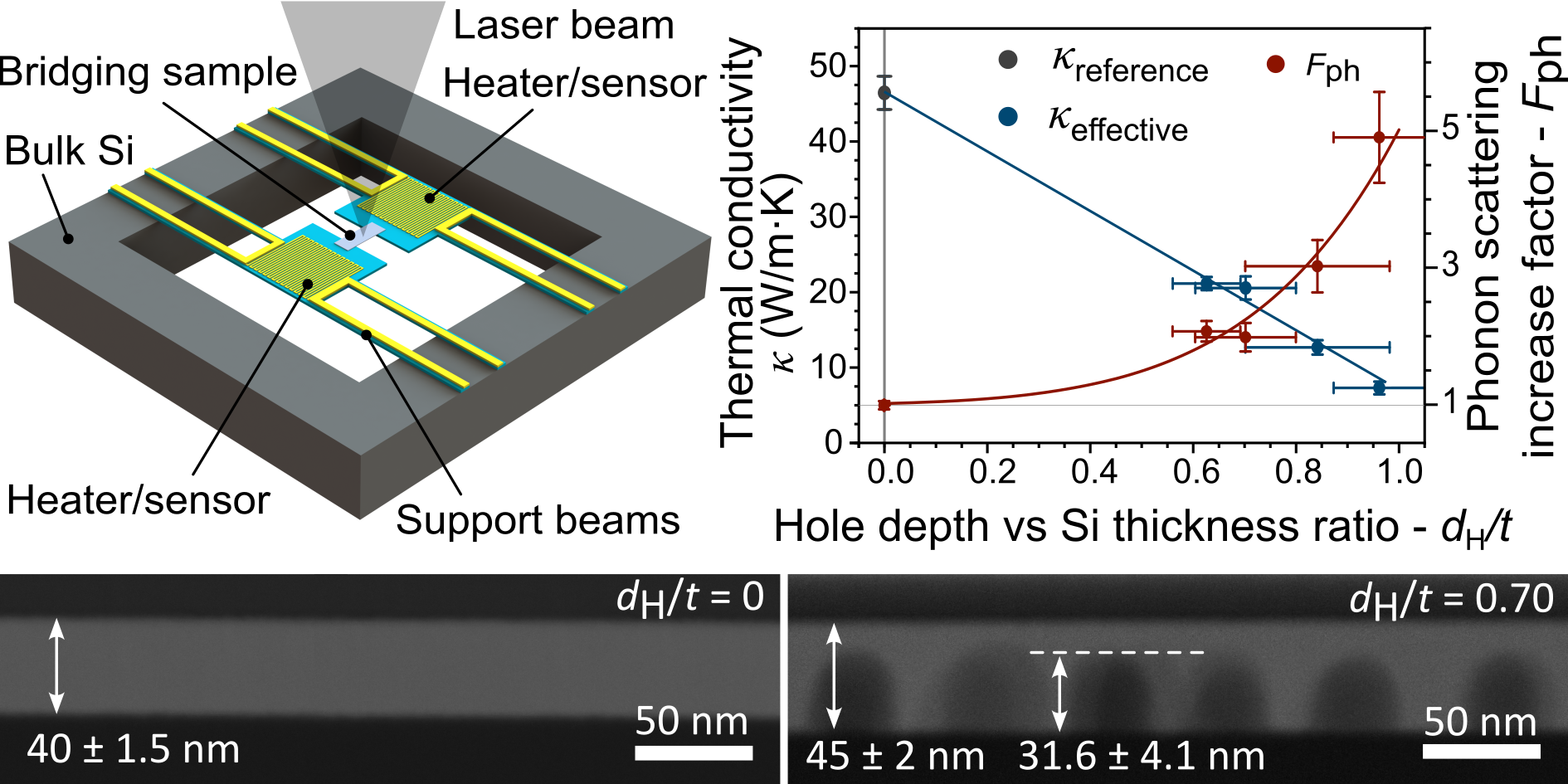}

\end{tocentry}

\begin{abstract}
Phononic silicon structures have emerged as an integrable and scalable nanosystem for tailoring thermal transport. However, their widespread adoption has been limited by their complex fabrication pathways. Alongside, the reliable characterization of thermal properties in suspended nanostructured films remains challenging, as thermal contact resistances often hinder the accuracy of measurements.

In this work, we demonstrate a clear and controllable reduction of thermal conductivity in nanopatterned silicon membranes. A block copolymer self-assembly approach is employed to fabricate nanoholed silicon films with a pitch of \qty{63}{nm} and hole diameters of \qty{35}{nm}. Additionally, we introduce an extension of the three-probe technique that enables robust, quantitative, and spatially resolved thermal conductivity measurements in complex thin-film systems, accounting for thermal contact artifacts.

The method is validated through measurements on unpatterned $\sim$\qty{40}{nm}-thick silicon thin films between 30 and \qty{350}{K}, yielding a room-temperature thermal conductivity of \qty{46.5}{W/m.K},\textcolor{markCol1}{ in good agreement with reported data.} Finally, we further show that controlled etching of the nanoholes provides a powerful means to tune thermal transport in the overall studied temperature range, \textcolor{markCol1}{showcasing hole etch depth as an effective parameter in phononic crystal engineering.} Specifically, a \textcolor{markCol1}{sixfold reduction in the effective} thermal conductivity is achieved, reaching \qty{7.3}{W/m.K} for fully etched-through membranes at room temperature.

\end{abstract}
\clearpage

\section*{Nomenclature}

\begin{table}[H]
\centering
\caption{List of magnitudes, subscripts, and superscripts used in the model.}
\begin{tabular}{ll}
\hline
\textbf{Magnitude} & \textbf{Description} \\
\hline
$Q$      & Heat flux [W] \\
$G$      & Conductance [W/K] \\
$\theta$ & Thermal difference with respect to the bath temperature [K] \\
$R$      & Resistance (electrical) [$\Omega$] \\
$\mathcal{R}$     & Resistance (thermal) [K/W] \\
$P$      & Applied power (dissipated) [W] \\
$\Psi$   & Absorbance [-] \\
$x$      & Longitudinal coordinate [m] \\
$y$      & Perpendicular coordinate [m] \\
$\Gamma$ & Beam size (Gaussian standard deviation) [m] \\
$W$      & Sample width [m] \\
$L$      & Sample length [m] \\
$I$      & Integrated intensity [W] \\
$\rho$   & Hole density [1/m$^{2}$] \\
$d$      & Hole depth [m] \\
$t$      & Thickness [m] \\
$\Phi$   & Diameter [m] \\
$\varrho$   & Porosity [-] \\
$V$      & Volume [m$^3$]\\
$v$      & Volume per unit area [m$^3/$m$^2$]\\
\hline
\textbf{Subscripts} & \textbf{Description} \\
\hline
$X_C$     & Contact \\
$X_{b}$   & Platform support beam \\
$X_{Br}$  & Bridge (sample + contacts) \\
$X_{NS}$  & Nanostructure \\
$X_{L}$   & Left side \\
$X_{R}$   & Right side \\
$X_{0}$   & Position of the laser \\
$X_{H}$   & Heater side \\
$X_{S}$   & Sensor side \\
$X_{SiN}$ & Silicon nitride platform \\
\hline
\textbf{Superscripts} & \textbf{Description} \\
\hline
${X^{\times}}$   & No heating case \\
$X^{\downarrow}$      & Laser heating case \\
$X^{\Omega}$ & Heat balance (laser and ohmic heating) case \\
\hline
\end{tabular}
\end{table}

\begin{table}[H]
\centering
\caption{List of abbreviation used in this work.}
\begin{tabular}{@{}ll}
NS    & Nanostructure\\
TCR   & Temperature Coefficient of Resistance\\
MEMS  & Micro Electro-Mechanical System\\
SMU   & Source Meter Unit\\
FEA   & Finite Element Analysis\\
TEM   & Transmission Electron Microscope\\
PCB   & Printed Circuit Board\\
BCP   & Block Copolymer\\
SOI   & Silicon On Insulator\\
BOX   & Buried Oxide Layer\\
BOE   & Buffered Oxide Etch\\
RIE   & Reactive Ion Etching\\
CPD   & Critical Point Drying\\
\end{tabular}
\end{table}

\section{Introduction}
Phononic engineering in semiconductors thin films is essential for controlling thermal transport, which is the backbone of scalable and silicon-compatible thermoelectric energy harvesting technologies\cite{Yanagisawa2020,Yanagisawa2024,Dimaggio2024,Ma2015}. At the nanoscale, thermal conductivity is no longer an intrinsic material property, as it becomes strongly dependent on geometry and on surface and interface scattering. Patterning thin films with periodic nanostructures has, therefore, emerged as a powerful strategy to tailor phonon transport beyond bulk limitations while maintaining compatibility with established microfabrication processes. However, achieving precise control over the pattern geometry while providing a scalable approach remains challenging. 

In this context, the self-assembly of block copolymers (BCP) offers a scalable and cost-effective route to define nanoscale patterns with high uniformity over large areas\cite{Hu2014}. When used as an etch mask, BCPs enable direct transfer of sub-lithographic features into silicon. This advantage makes BCPs particularly attractive for phonon engineering applications. Yet, the integration of BCP masks into silicon microdevice fabrication workflows and their direct use as etch masks has been vaguely explored in the context of thermal transport studies, with approaches often requiring intermediate masks that complicate the process\cite{Tang2010,Lim2016}.

Another challenge is the accurate experimental quantification of the thermal conductivity in such patterned thin films. Due to the intrinsic randomness of the resulting contact area of transferred samples, and thus of the contact thermal resistance, the use of the well-established bridge method \cite{Shi2003} can lead to large measurement errors. These errors arise from the fact that the method is incapable of separating the contribution of the thermal contact resistance from the conductance of the sample, or to address sample spatial inhomogeneities. Hence, the three-probe method has recently emerged as a powerful technique to address these challenges, allowing local measurements of thermal resistance\cite{Huang2024}. \textcolor{markCol1}{While} previous implementations have relied mainly on electron-beam-based heating\cite{Lui2014}, the extension of this approach to optical excitation would significantly broaden its applicability and experimental flexibility.

In this work, we combine block copolymer–based nano-patterning with laser-based three-probe thermal measurements to investigate heat transport in patterned silicon thin films. The self-assembly of BCPs is employed to define periodic hole arrays, which are directly used as etch masks to produce well-controlled nanostructures with tunable hole depths. The films are then transferred to 2-terminal microcalorimeter devices\cite{Swinkels2015,Arya2025}. Subsequently, using a focused laser beam as an additional thermal probe, we extend the use of the three-probe method--originally conceived for 1D samples--to thin films. A study of the range of application for a 2D sample using finite element analysis (FEA) and a measurement of a pristine (unpatterned) silicon reference film validate the approach. Next, the fabricated films are morphologically inspected as a function of the etching time, showing accurate control of the etch depth. Finally, the thermal properties of the patterned films are evaluated and discussed in relation to their morphology.

\section{Materials and methods}
\label{sec:Expermental}

\subsection*{Device fabrication}
Devices featuring micro-suspended platforms were used for the thermal conductance evaluation of the samples. Each of these micro-platforms is thermally insulated from the bulk by long silicon nitride beams (with a total thermal conductance of $\sim$\qty{100}{nW/K}) and features a Pt micro-resistor ($\sim$\qty{20}{k \ohm}) which can be used both as heater and as thermometer owed to the Joule heating and the large temperature coefficient of resistance (TCR) of Pt. The fabrication process of such micro-suspended platform devices is detailed elsewhere\cite{Arya2025}.

\subsection*{Electro-thermal measurements}
Electrical measurements were performed using a Keithley 2635B featuring two independent Source-Meter Units (SMU), allowing simultaneous resistor biasing and readouts of heater and sensor. Resistance measurements were always extracted from fits of I-V curves to avoid voltage offset artifacts. During all measurements, the devices were loaded onto a sealed temperature controlled stage within an ARS DE200 cryostat chamber at a vacuum pressure of \qtyrange{1e-5}{1e-6}{mbar}. The absolute TCR (defined as $\partial R/\partial T$) of each resistor was calibrated by tracking the resistance change of the devices over the measured temperature range. A \qty{473}{nm} laser beam (C-Flex from Hubner photonics) was focused on the sample and used as the heating source across the thin film which is required for the sample characterization without thermal contact resistance contributions (see the description of the three-probe method at \autoref{sec:3ProbeMethod}). The spot size of the laser beam was estimated to be \qty{600}{nm} using the definition $1/e^2$ with a numerical aperture NA~=~0.5 of the objective (Mitutoyo 100x with a working distance of \qty{12.06}{mm}). The position of the laser beam was controlled using a custom-made 3-axis optical set-up motorized with feedback controlled stepper stages (PI-C663 Mercury-Step). 

\subsection*{Nanomembrane fabrication}
\begin{figure}[!htb]
\begin{center}
    \includegraphics[width=1.0\textwidth]{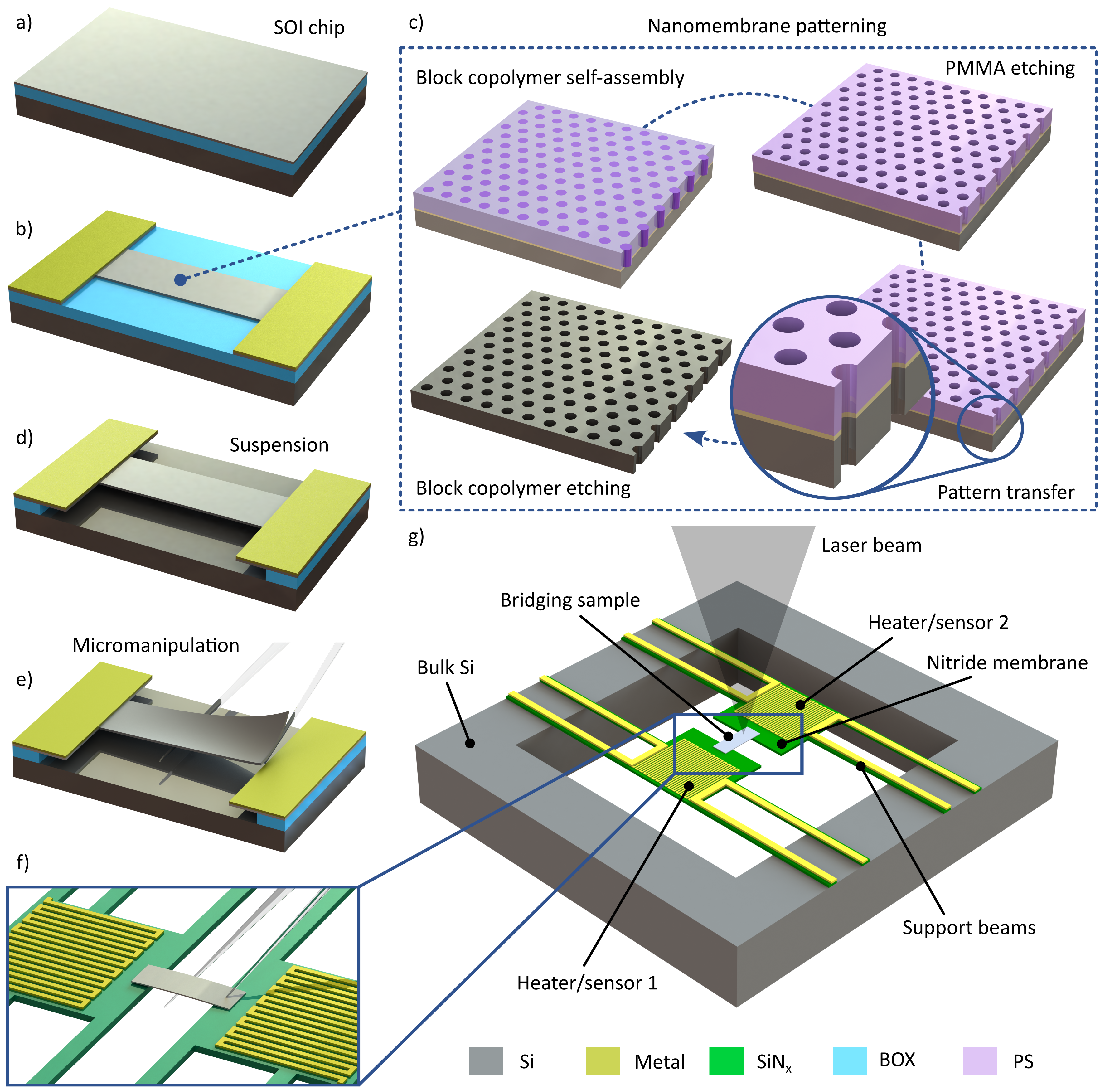}
      \caption{a)~Initial SOI chip. b)~Photolithography\textcolor{markCol1}{, evaporation} and lift-off to define the metal frame, followed by photolithography and RIE to define the membranes layout. c)~Nanostructuration of the Si film with the block copolymer pattern transfer method. d)~BOE wet etching of the BOX with critical point drying to suspend the membranes. e)~Cutting of the membrane with two glass tips using micro-positioners. f)~Transferring of the membrane onto the bridge device. g)~Zoom out of the bridge device illustrating the support beams and the incident focused laser beam.}
      \label{fig:Fabrication}
\end{center}
\end{figure}  

The membrane fabrication starts with a silicon on insulator (SOI) wafer composed of three layers: a \qty{725}{\micro m} bulk Si layer, a \qty{400}{nm} buried oxide layer (BOX), and a \qty{50}{nm} Si layer on top with a resistivity of \qtyrange{8}{12}{\ohm.cm} (\autoref{fig:Fabrication}a). The wafer has a (100) crystal orientation. After substrate cleaning, Cr/Au metal frames (10 and \qty{60}{nm} thick, respectively) with gaps ranging from 12.5 to \qty{20}{\micro m} are defined through metal evaporation, photolithography, and lift-off. \textcolor{markCol1}{The layout of the membranes are defined within them (membrane widths of \qty{5.5}{\micro m}) using photolithography and RIE} (see \autoref{fig:Fabrication}b). The metal frames serve to facilitate the membrane transfer later on. Afterwards, the exposed \qty{50}{nm} thick Si layer is nanostructured using block copolymers (BCP~--~PS-b-PMMA in this work) and pattern transfer (see \autoref{fig:Fabrication}c). For this purpose, PS-b-PMMA films with a nominal cylindrical pattern of period $\sim$ \qty{57}{nm} are deposited and self-assembled with controlled pattern orientation thanks to the use of a brush layer (i.e. a PS-r-PMMA random copolymer monolayer grafted with a 74\% PS content and an \qty{8.4}{kg/mol} molecular weight). The \qty{57}{nm} period is set by blending cylindrical BCPs with \qty{36}{nm} and \qty{80}{nm} periods (molecular weights of 60.8 and \qty{219.6}{kg/mol}, and PS contents of 69.08\% and 70.4\%, respectively) in a 50:50 ratio. Base BCPs were supplied by Arkema S.A. BCP spinning conditions are \qty{1500}{rpm} for \qty{30}{s} using a 1\% wt solution of PS-b-PMMA in propylene glycol methyl ether acetate (PGMEA) to obtain \qty{35}{nm}-thick PS-b-PMMA films. Annealing conditions to graft the PS-r-PMMA and self-assemble the PS-b-PMMA are \qty{230}{\degree}C for \qty{5}{min} and \qty{230}{\degree}C for \qty{10}{min}, respectively. Later, PMMA is etched by exposing the samples to UV light (\qty{18.5}{mW/cm^2} irradiation intensity, $\lambda$ = \qty{254}{nm}) for \qty{230}{s} to break the PS-PMMA covalent bonds and induce chain scission in PMMA, followed by the dissolution of PMMA in acetic acid. Then, the PS-b-PMMA pattern is transferred into the underlying Si layer using a mixed Bosch reactive ion etching (RIE). Finally, the remaining PS mask is etched using oxygen plasma. After nanostructuring the Si layer, the membranes are defined within the metal frames (membrane widths of \qty{5.5}{\micro m}) using photolithography and RIE. The fabrication ends with the release of the membrane. For this purpose, buffered oxide etch (BOE) is used to remove the BOX layer beneath the membranes (see \autoref{fig:Fabrication}d). The samples are finally dried using critical point drying (CPD) to avoid membrane collapse. 

\subsection*{Membrane transfer}
The membranes are transferred onto the bridge devices using an optical microscope featuring a pair of hydraulically-actuated micro-manipulator stages (Narishige MMO-4). A pair of $\sim$ \qty{200}{nm} sharp glass needles are used for the transfer of the film, fabricated with a Narishige PC-100 puller. Firstly, membranes are detached from the chip by pressing with the tips at the edges where the membranes connect to the metal frame (see \autoref{fig:Fabrication}e). During this process, the second tip is placed beneath the center of the membrane to prevent it from adhering to the substrate after one side has been cut. Afterwards, the membranes are placed onto the two nitride platforms of the \textcolor{markCol1}{bridge device} (see \autoref{fig:Fabrication}f). To ensure good contact, membranes are pushed down with the tips against the platforms until the fringe patterns of the nitride underneath are clearly visible on the membrane. Finally, the chips are silver-pasted and wire-bonded to a custom highly conductive copper core printed circuit board (PCB) that also clamps a diode temperature sensor (Lakeshore DT-670).

\subsection*{Transmission Electron Microscopy}
The structural and morphological characterization of membranes was conducted on twin samples adjacent to each of the studied ones. For each of the membranes used in the thermal analysis, another membrane was transferred onto silicon nitride on silicon substrates. These twin membranes were subsequently prepared as electron-transparent lamellas for HR-TEM imaging in cross-sectional geometry. To this end, a Zeiss Crossbeam 540 operated at acceleration voltages of 30, 5, and \qty{2}{kV} was used, subsequent to the deposition of carbon and platinum protective layers. The final thinning was performed at ±~\qty{0.5}{\degree} incident angle at \qty{30}{kV} followed by low-kV polishing at ±~\qty{8}{\degree} to minimize ion beam-induced damage at the surface of the lamellas. HR-TEM images of the samples were taken on a JEOL JEM F200 cFEG operating at \qty{200}{keV} and equipped with an EMSIS XAROSA CMOS camera and the JEOL JED-2300 analysis station for STEM. Atomically resolved HAADF-STEM images were recorded on a FEI Titan Themis operated at \qty{300}{kV} and equipped with a probe CEOS DCOR spherical aberration corrector. The probe semiconvergence angle was set to \qty{18}{mrad} in combination with an annular semidetection angle of \qtyrange{66}{200}{mrad} for the annular dark field detector.

\subsection*{Thermal characterization approach}
The thermal characterization of the samples was carried out as follows. Firstly, thermal conductance measurements were performed with the standard bridge method\cite{Swinkels2015,Arya2025} over the whole temperature range (\qtyrange{20}{350}{K}). This method measures the electrical resistance increase of both the heated platform and the opposite platform acting as a sensor. The same dataset was also used to calibrate the absolute TCR ($\partial R/\partial T$), which allows to extract the temperature rise as a function of the dissipated power (\autoref{fig:MethodSketch}a). Secondly, a three-probe technique was applied at \qty{300}{K} \cite{Lui2014,Huang2024} (see detailed description in \autoref{sec:3ProbeMethod}), to separate the thermal conductance of the sample from the thermal contact resistance contributions ($\mathcal{R}_C$). The obtained $\mathcal{R}_C$ at \qty{300}{K} was assumed constant over the whole temperature range. With \textcolor{markCol1}{this} assumption, the underestimated conductances obtained with the bridge method were corrected accordingly\textcolor{markCol1}{, but the uncertainty associated with the correction increases as the extrapolation departs from the measurement temperature, due to different temperature behaviors between the sample itself and the contact resistance.}


\section{Thermal three-probe method} \label{sec:3ProbeMethod}

Prior to the application of the three-probe method, a standard thermal bridge experiment must be performed~\cite{Swinkels2015,Arya2025}, as represented in \autoref{fig:MethodSketch}a. This preliminary step measures the resistances $R_{\mathrm{side}}^{\times}$, the absolute TCR $(\partial R/\partial T)_{\mathrm{side}}^{\times}$, the temperature rise of the sensor side as a function of the heater power $(\partial \theta_S/\partial P_{Ht})^{\times}$, and the beam conductance $G_{b,\mathrm{side}}^{\times} = ({\mathcal{R}_{b,\mathrm{side}}})^{-1}$ of each microheater as a function of temperature. Note that variables measured in this phase are indicated by the superscript $\times$.

\begin{figure}[!tb]
\begin{center}
    \includegraphics[width=1.0\textwidth]{Figure_MethodSketch}
      \caption{Sketches describing the different steps of the method. \textcolor{markCol1}{Red arrows represent dissipated powers (Joule or laser) whereas black arrows represent heat flux across the system.} a)~Standard bridge method measurement that determines $(\partial \theta_S/\partial P_{Ht})^{\times}$. b)~Heating with laser step for determining $\theta_S^{\downarrow}$, $Q_S^{\downarrow}$, and the absorption $\psi$. c)~Balancing step achieved by inserting $P_{Ht}^{\Omega}$. The shaded lines illustrate how the temperature in the heater section is balanced by the superposition of the two heat sources. In these conditions $\theta_S^{\Omega}$ and $\theta_{Ht}^{\Omega}$ are measured to determine $\theta_0^{\downarrow}$.}
      \label{fig:MethodSketch}
\end{center}
\end{figure}

The implementation of the three-probe method follows two measurement phases, which are repeated at each sample position and at several laser power levels. Firstly, at the laser heating phase (denoted with the $\downarrow$ superscript), the laser is focused at a position $x_0$ on the sample, causing both platforms to heat up (\autoref{fig:MethodSketch}b). A resistance measurement with negligible electrically dissipated power ($<$\qty{150}{nW}) is performed to obtain the resistance increase with respect to the case without laser heating $ \Delta R_{\mathrm{side}}^{\downarrow} = R_{\mathrm{side}}^{\downarrow} - R_{\mathrm{side}}^{\times} $. Using the calibrated absolute TCR $(\partial R/\partial T)_{\mathrm{side}}^{\times}$, the temperature increase $\theta_{\mathrm{side}}^{\downarrow}$ and the heat flow $Q_{\mathrm{side}}^{\downarrow}$ are obtained:
\begin{equation}
Q_{\mathrm{side}}^{\downarrow}
= G_{b,\mathrm{side}}^{\times} \, \theta_{\mathrm{side}}^{\downarrow}
= G_{b,\mathrm{side}}^{\times} \, \Delta R_{\mathrm{side}}^{\downarrow}
\left( \frac{\partial T}{\partial R} \right)_{\mathrm{side}}^{\times}.
\end{equation}

Secondly, the ohmic balance phase is conducted (denoted with the $\Omega$ superscript). Now, one of the resistors is biased so that it dissipates a power $P_{\mathrm{Ht}}^{\Omega}$. This power is adjusted so that the temperature of the heater equals the temperature at the laser spot $\theta_{\mathrm{Ht}}^{\Omega} = \theta_0^{\Omega}$ (\autoref{fig:MethodSketch}c)~\cite{Huang2024}. This condition is reached when the heat flowing into the heater side under laser illumination, $Q_{\mathrm{Ht}}^{\downarrow}$, is balanced by an equal flow $Q_S^{\Omega}$ in the opposite direction, as depicted in \autoref{fig:MethodSketch}c:
\begin{equation}
G_{b,\mathrm{Ht}}^{\times} \, \theta_{\mathrm{Ht}}^{\downarrow}
= Q_{\mathrm{Ht}}^{\downarrow}
= Q_S^{\Omega}
= G_{b,S}^{\times} \, \theta_S^{\Omega}.
\label{eq:HeatBalance}
\end{equation}

The required power $P_{\mathrm{Ht}}^{\Omega}$ is calculated using the calibrated curve from the bridge experiment $(\partial \theta_S/\partial P_{Ht})^{\times}$:

\begin{equation}
P_{\mathrm{Ht}}^{\Omega}
= \frac{G_{b,\mathrm{Ht}}^{\times}}{G_{b,S}^{\times}}
\frac{\theta_{\mathrm{Ht}}^{\downarrow}}
     { (\partial \theta_S/\partial P_{Ht})^{\times} }.
\end{equation}

While dissipating $P_{\mathrm{Ht}}^{\Omega}$, the \textcolor{markCol1}{electrical} resistance change at both platforms is measured again as $ \Delta R_{\mathrm{side}}^{\Omega} = R_{\mathrm{side}}^{\Omega} - R_{\mathrm{side}}^{\times} $. The corresponding temperatures and heat fluxes are calculated using $(\partial T/\partial R)_{\mathrm{side}}^{\times}$ and $G_{b,\mathrm{side}}^{\times}$. Noteworthy, since additional power is being dissipated, the peak temperature in the system is higher, i.e. $\theta_0^{\Omega} > \theta_0^{\downarrow}$. Therefore, to extract useful information from the balancing data, the original temperature at the laser spot size $\theta_0^{\downarrow}$ needs to be calculated. Redefining the total thermal resistance of the bridging sample as the sum of two contributions on each side of the laser position 
$\mathcal{R}_{\mathrm{Br}} = \mathcal{R}_{\mathrm{Br},L}(x) + \mathcal{R}_{\mathrm{Br},R}(x)
= [G_{\mathrm{Br},L}(x)]^{-1} + [G_{\mathrm{Br},R}(x)]^{-1} $
and assuming that the conductances remain unchanged under the small temperature rise (in our experiment, limited to a maximum of \qtyrange{30}{50}{K}), the ratio of heat flow for the laser heating and ohmic balance cases allows the calculation of $\theta_0^{\downarrow}$ as:

\begin{equation}
\theta_0^{\downarrow}
= \theta_S^{\downarrow}
+ \frac{G_{b,S}^{\times}\, \theta_S^{\downarrow}}{P_0 \psi}\,
\left( \theta_{\mathrm{Ht}}^{\Omega} - \theta_S^{\Omega} \right).
\end{equation}
where one can assume $Q_S^{\downarrow} = G_{b,S}^{\times} \theta_S^{\downarrow}$ and $\theta_{\mathrm{Ht}}^{\Omega} = \theta_0^{\Omega}$ at the balance condition. Since the procedure can be carried out using either of the two platforms as the heater, two measurements of $\theta_0^{\downarrow}$ can be obtained at a given laser position $x_0$. The variance of these two results indicates the accuracy of the thermal balance achieved. In addition, at each laser position, this experiment is repeated with several laser powers, so that a trend can be derived.

\textcolor{markCol1}{Noteworthy, in a similar way to the bridge method, in the three-probe approach the range of application is also limited by the relative thermal resistance of the sample with respect to the contacts and the suspended platform beam ones. For an optimal resolution, bridge and beam resistances need to be matched \cite{Villalba2016}. In addition, it is clear that the contact and sample resistances need to be of the same order of magnitude. For too large $\mathcal{R}_{C}$, the change in temperature along the sample becomes small and cannot be resolved, whereas if $\mathcal{R}_{C}$ is relatively small, the bridge method itself suffices.}

\section{Validity of the method on 2D membranes}
While the model assumes a \textcolor{markCol1}{linear} heat source at the sample \textcolor{markCol1}{that would force 1D conduction}, the actual power deposited by the laser beam follows a 2D Gaussian distribution of the type:
\textcolor{markCol1}{
\begin{equation}
p(x,y,\Gamma,x_0,y_0) = A \exp{ \left( - \frac{(x - x_0)^2+(y - y_0)^2}{2 \Gamma^2} \right) }
\label{eq:GaussPower}
\end{equation}
}
where, $A = P_0/(2 \pi \Gamma^2)$ such that the integral of $p(x,y)$ is $P_0$. 
\textcolor{markCol1}{Therefore, a careful evaluation of the sample dimensions with respect to the beam size must be performed prior to any measurement. Any fraction of power dissipated directly in the silicon nitride skips the lumped contact resistance $\mathcal{R}_{C}$ of the 1D model (\autoref{fig:MethodSketch}) and, therefore, introduces an error. This error is expected to be tightly linked to the ratio of the power dissipated in the nitride platforms over the one on the suspended sample, which can be calculated as $P_{SiN}/P_{NS} = (I_{SiN} \Psi_{SiN})/(I_{NS} \Psi_{NS} )$, where $\Psi_{SiN}/\Psi_{NS}$ is the ratio of the materials' absorptions and the $I_{SiN}$ and $I_{NS}$ are the integrated powers of the Gaussian beam 2D function $p(x,y,\Gamma,x_0,y_0)$ at the silicon nitride area and the suspended membrane areas, respectively:}
\textcolor{markCol1}{
\begin{equation}
    I_{SiN} = \int\limits_{L/2}^{\infty} \int\limits_{-\infty}^{\infty} p(x, y, \Gamma, x_0, 0)\, dy \, dx,
    \qquad
    I_{NS} = \int\limits_{-L/2}^{L/2} \int\limits_{-W/2}^{W/2} p(x, y, \Gamma, x_0, 0)\, dy \, dx
    \label{eq:GaussIntegrals}
\end{equation}
}

\begin{figure}[!tb]
\begin{center}
    \includegraphics[width=0.55\textwidth]{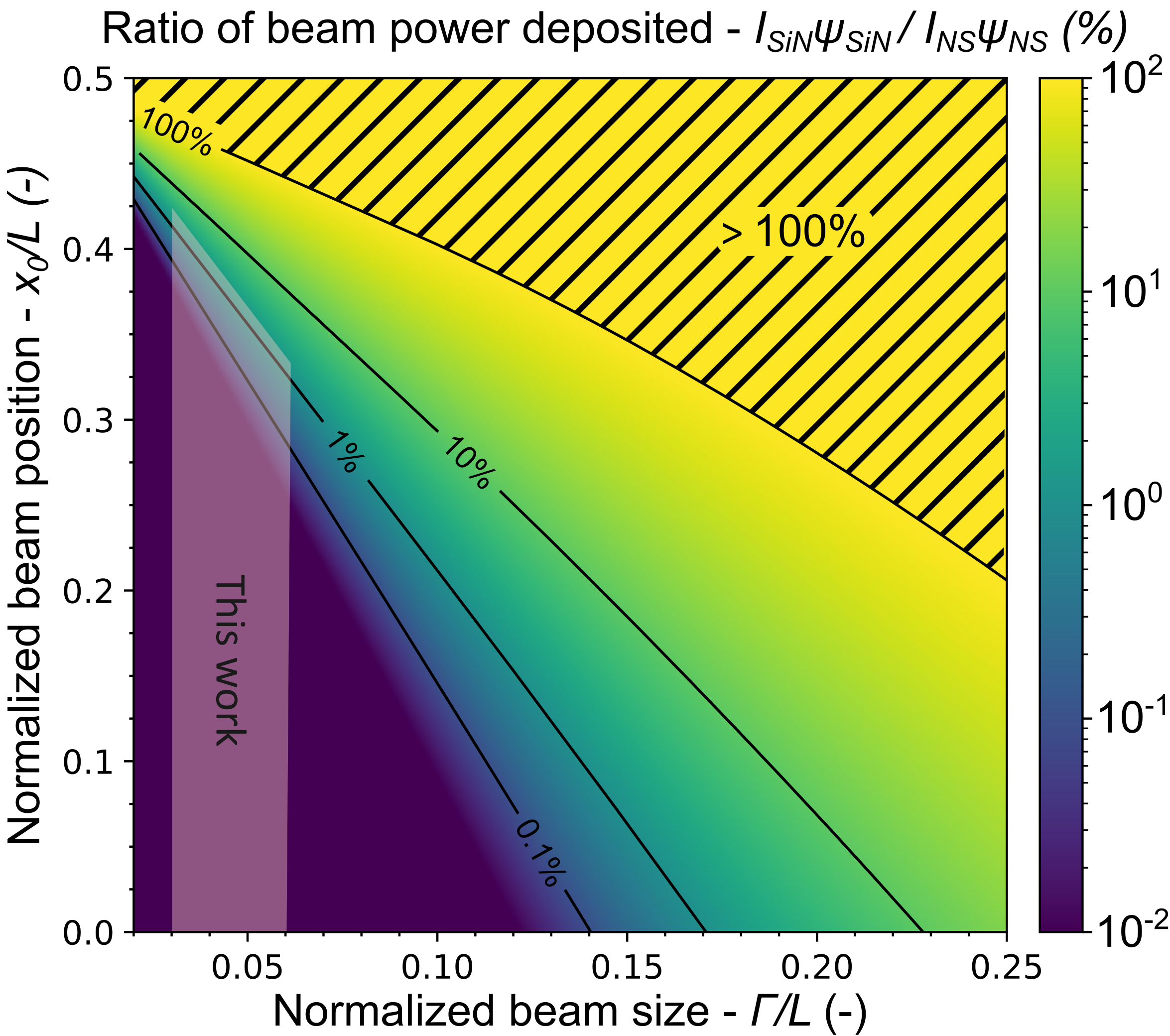}
      \caption{ \textcolor{markCol1}{Calculated ratio of beam power dissipated on the nitride with respect to the one on the suspended nanostructure for $W/L = 0.5$. Ratio is scaled by the different absorption of both materials ($\Psi_{SiN}/\Psi_{NS} = 5$). The highlighted area shows the range used in this study.} }
      \label{fig:BeamSizeError}
\end{center}
\end{figure}  

\textcolor{markCol1}{In this work, a conservative value of 5 for $\Psi_{SiN}/\Psi_{NS}$ is found experimentally (see \autoref{fig:3ProbeResults}c). However, this value is strongly dependent on the thickness and nature of the nitride and sample materials and must be considered case-wise. \autoref{fig:BeamSizeError} shows the estimation obtained. A clear monotonic trend appears. The larger the relative beam size ($\Gamma/L$), the lower the normalized longitudinal beam position range ($x_0/L$) where the error is contained below reasonable levels. As a general rule of thumb, sample length must be at least 10 times larger than the beam size in order to reliably scan at least a 50 to 60\% of the total sample length. The shaded area represents the range covered by the samples studied in this work. As it can be observed, errors derived from this issue are expected to be below 1\%.}

\begin{figure}[!hptb]
\begin{center}
    \includegraphics[width=1.0\textwidth]{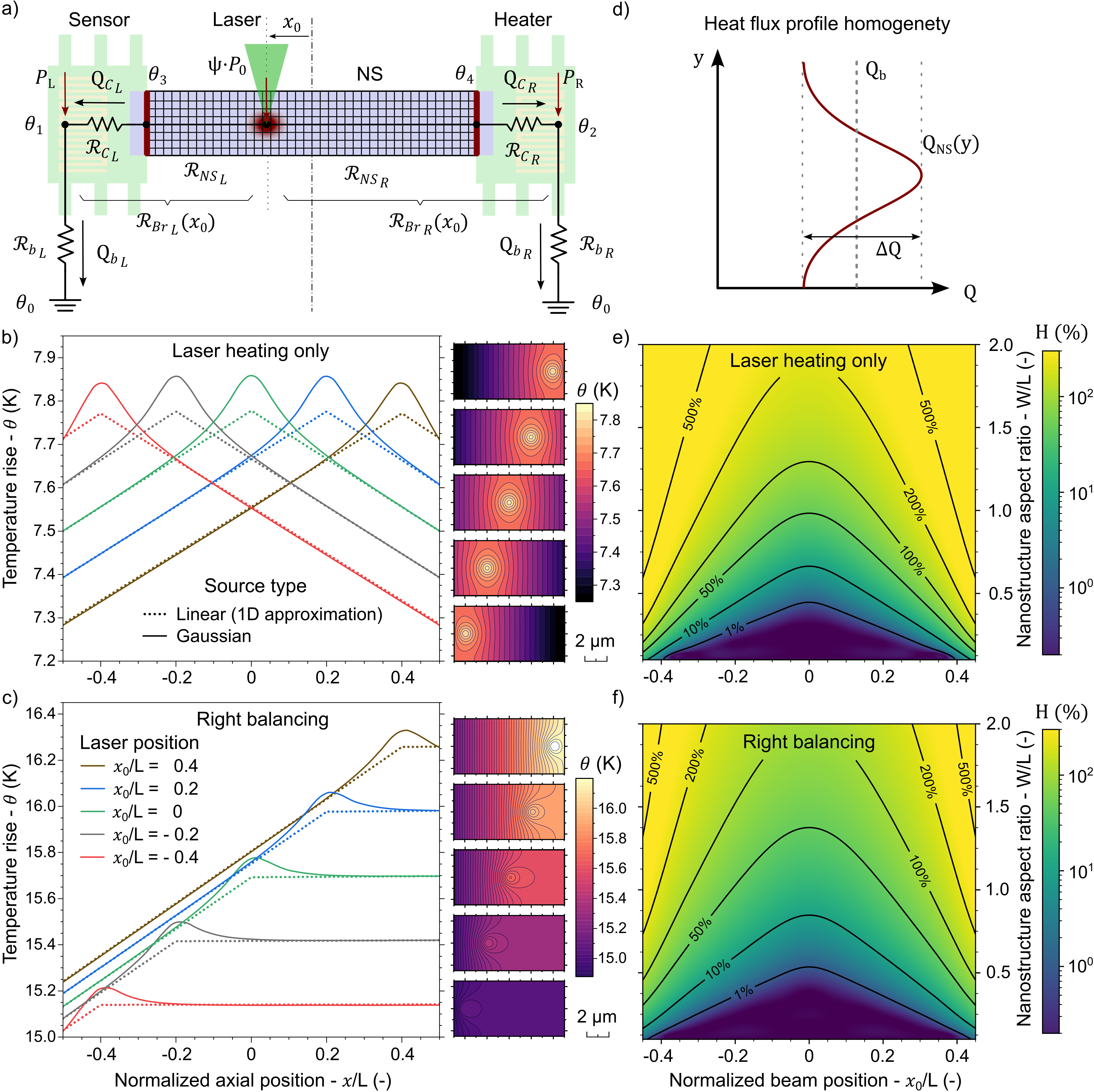}
      \caption{
      \textcolor{markCol1}{
      a)~Not to scale sketch of the FEA 2D model, with a discretized 2D nanostructure and a lumped resistive model for the contact resistance and the platform beams. Vertical red edges indicate where the heat inhomogeneity parameter is evaluated. 
      b)-c)~Temperature rise produced by a Gaussian heating source ($\Gamma$ = \qty{0.3}{\micro m}) at different $x_0/L$ with $W/L = 0.5$ under the beam heating only~(b) and right-side balancing~(c) conditions. The left charts illustrate the temperature profile cuts (solid) at $y = 0$ from the temperature maps illustrated in the right-side while the dashed lines represent the ideal profile arising from a linear source constraining heat dissipation to the longitudinal axis (1D approximation).
      d)~Sketch illustrating the definition of the heat flux inhomogeneity $H$.
      e-f)~Maps of $H$ as a function of the sample aspect ratio $W/L$ and the beam position $x_0/L$ at the beam heating only~(e) and right-side balancing~(f) conditions.}
      }
      \label{fig:2D_FEM}
\end{center}
\end{figure}

\textcolor{markCol1}{Another consequence of a Gaussian power beam is} that the temperature at the sample does not have a sharp maximum, but rather a smoothed one. This can raise concerns about the validity of the model used, as it is based on a series of lumped thermal resistances, i.e. it assumes one-dimensional heat transport. To validate this approach, a 2D FEA was performed. \autoref{fig:2D_FEM} provides a schematic of the simulated 2D sample with a suspended width-to-length aspect ratio ($W/L$) of 0.5 and a length significantly larger than the spot size (i.e. membrane length of \qty{10}{\micro m}, width of \qty{5}{\micro m}, and spot size with \textcolor{markCol1}{$\Gamma = $\qty{300}{nm})}. This aspect ratio is representative of the samples studied in this work (see \autoref{tab:StrucuturalResults}). In this study, this Gaussian model is compared with the one obtained from a linear source, which constrains heat dissipation to the longitudinal axis. This yields the same thermal profile of the 1D approximation assumed in the lumped thermal model proposed in \autoref{fig:MethodSketch}b.  

\textcolor{markCol1}{The temperature maps illustrated in \autoref{fig:2D_FEM}b corroborate the estimation of \autoref{fig:BeamSizeError}.} For positions of the heating beam at \textcolor{markCol1}{$x_0/L = \pm$\num{0.4}}, the Gaussian tail of the beam fits reasonably well within the suspended zone of the sample. \textcolor{markCol1}{Hence, no noticeable} heat is directly deposited on the microdevice nitride platforms, i.e. outside of the simulated suspended area. In this condition, the resulting temperature profile, represented with solid lines in the left graph of \autoref{fig:2D_FEM}b, matches the 1D approximation assumed by the model (represented with the dotted lines) already at a distance of \qty{1}{\micro m} from the beam position and all the way towards the contacts (\textcolor{markCol1}{$x_0/L = \pm$\num{0.5}}). Likewise, during the "thermal balancing" phase (\autoref{fig:2D_FEM}c), the thermal gradient estimated with the 1D approximation--and therefore the heat fluxes--still matches the actual 2D temperature gradient away of the beam heated area. 

Noteworthy, this 1D approximation (the dotted lines in \autoref{fig:2D_FEM}b~and~c) yields a single virtual beam-source temperature $\theta_0^\downarrow$ which differs from the actual maximum value. Since the heat balance still holds \textcolor{markCol1}{$\sim3 \Gamma$} away from the beam source, both at the \textcolor{markCol1}{"laser heating only"} (\autoref{fig:2D_FEM}b) and the "thermal balancing" (\autoref{fig:2D_FEM}c) phases, $\theta_0^\downarrow$ corresponds to the spatially integrated average temperature generated by the Gaussian source. This single temperature value is useful for computing the cumulated thermal resistance ${\mathcal{R}_{Br}}_{side} (x)$ using the 1D approximation (lumped thermal resistance model described in \autoref{fig:MethodSketch}b). \textcolor{markCol1}{However, as $W/L$ increases, this approximation is more prone to fail if spatial inhomogeneities in the contacts are present. Hence, a magnitude to describe the heat flux inhomogeneity at the boundaries of the sample (red edges in \autoref{fig:2D_FEM}a) is defined as:
\begin{equation}
    H = \sqrt{\left(\frac{\Delta Q}{Q_b}\right)_L^2 + \left(\frac{\Delta Q}{Q_b}\right)_R^2}
    \label{eq:HeatInhomogenety}
\end{equation}
with $\Delta Q = \text{range}( Q_x (x=\pm L/2) )$, where $\text{range}(Q) = \max(Q)  – \min(Q)$, as sketched in \autoref{fig:2D_FEM}d. In general, $H>100\%$ represents conditions where part of the heat is locally inverted across the vertical boundary between the suspended nanostructure and the microplatform (even when the total integrated one still presents a net rate that exits the nanostructure). These conditions are highly undesired, as they can break the thermal resistance model, and therefore can yield nonphysical results. As it can be appreciated in the maps of $H$ as a function of $W/L$ and $x_0/L$ (\autoref{fig:2D_FEM}e~and~f), for $W/L\le0.5$, a wide range of the nanostructure suspended length ($\sim$70\%) can be scanned with reasonable homogeneity. Overall, $H$ improves during balance conditions as the heated platform  helps the homogenization of the temperature profile when the beam is close to that platform (see \autoref{fig:2D_FEM}c~and~f).}

Therefore, this FEA proves that, for the simulated aspect ratio of $W/L\le0.5$, despite the 2D thermal gradient generated at the sample, the use of the 1D heat flow approximation is still accurate and thus justified. This interestingly large aspect ratio range in which 1D approximations work accurately enables the use of wider heat source beams, like those produced by lasers, rather than precise but inconvenient electron beams\cite{Lui2014,Huang2024}. This also avoids the need to experimentally account for the beam size itself, as long as it is completely contained within the suspended sample \textcolor{markCol1}{as described by \autoref{fig:BeamSizeError}}. The latter condition is experimentally simple to check, as the absorption of the laser beam drastically changes when the laser is focused over a part of the membrane that lies on the suspended platforms (see \autoref{fig:3ProbeResults}c).

\section{Results and discussion}

\begin{figure}[!htb]
\begin{center}
    \includegraphics[width=0.5\textwidth]{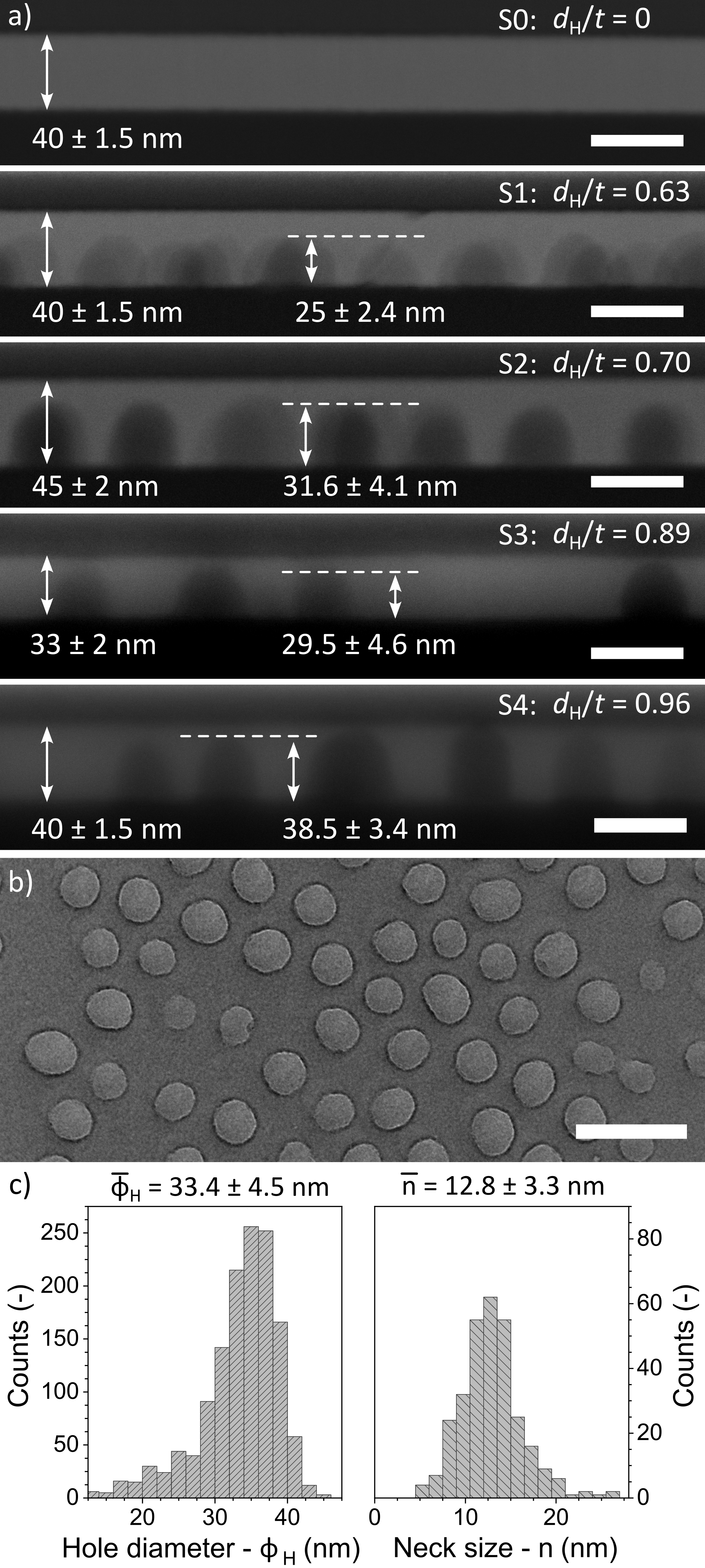}
      \caption{Study of $\sim$\qty{40}{nm} thick nano-patterned Si thin films with varying depth etch ratio $d_H/t$. a)~STEM cross section images of the samples illustrating the measured etch ratio $d_H/t$. All scale bars are \qty{50}{nm}. \textcolor{markCol1}{b)~Example of top view TEM image of sample S3 used for the study of hole density distribution. Scale bar is \qty{100}{nm}. c)~Histograms showing the hole diameter (left) and neck distance (right) distributions.}} 
      \label{fig:STEM}
\end{center}
\end{figure}

In this study, 4 samples with different hole depths were investigated, along with a reference pristine membrane without patterned holes, denoted as S0. The morphological analysis of the films was carried out on samples adjacent to those transferred onto the bridge microdevices. These twin samples were used to prepare lamellas, which were subsequently imaged by Scanning Transmission Electron Microscopy (STEM) to precisely measure the average depth, thicknesses, diameter, and density of the holes (see \autoref{fig:STEM}a). Since the average thickness of the films ranged between 33 and \qty{45}{nm}, the hole depth to Si thickness ratio ($d_H/t$) parameter was defined to compare the results throughout the different samples. Moreover, top-view TEM image analysis allowed to calculate an average hole size of \qty{35(3)}{nm} (\autoref{fig:STEM}b). A summary of all measured samples and their resulting parameters can be found in \autoref{tab:StrucuturalResults}. \textcolor{markCol1}{The statistical study of the average hole diameter and neck sizes (\autoref{fig:STEM}c) yielded \qty{33.4(4.5)}{nm} and \qty{12.8(3.3)}{nm}, respectively.} Notably, since the BCP patterning process was identical for all samples, the hole density was consistent across all samples, and the average value of \qty{227(8)}{\micro m^{-2}} was used to derive geometrical parameters.  

\begin{table}[!tb]
\centering
\caption{Summary of the morphological characterization of each sample studied.}
\label{tab:StrucuturalResults}
\resizebox{\textwidth}{!}{
\begin{tabular}{ccccccccc}
Sample & Hole density                    & RIE time          & Hole depth        & Thickness       & Etch ratio & Length & Width & \textcolor{markCol1}{Geom. factor}\\
       & $\rho_H$ (\unit{\micro m^{-2}}) & $\tau$ (\unit{s}) & $d_H$ (\unit{nm}) & $t$ (\unit{nm}) & $d_H/t$ (-) & $L$ (\unit{\micro m}) & $W$ (\unit{\micro m}) & \textcolor{markCol1}{$F_G$ (-)}\\
\hline
S0  & 0                             & 0  & 0               & \num{40(1.5)} & 0.0        & \num{8.68} & \num{3.69} & \textcolor{markCol1}{\num{1}} \\
\hline
S1  &                                      & 12 & \num{25.0(2.4)} & \num{40(1.5)} & \num{0.63(0.07)} & \num{9.44} & \num{4.33} & \textcolor{markCol1}{\num{1.12(0.07)}}\\
S2  &                                      & 15 & \num{31.6(4.1)} & \num{45(2.0)} & \num{0.70(0.10)} & \num{9.93} & \num{3.60} & \textcolor{markCol1}{\num{1.13(0.08)}}\\
S3  &                                      & 14 & \num{29.5(4.6)} & \num{35(2.0)} & \num{0.84(0.14)} & \num{10.2} & \num{5.645} & \textcolor{markCol1}{\num{1.21(0.12)}}\\
S4  & \multirow{-4}{*}{\num{227(8)}}   & 20 & \num{38.5(3.4)} & \num{40(1.5)} & \num{0.96(0.09)} & \num{9.99} & \num{2.73} & \textcolor{markCol1}{\num{1.30(0.09)}}\\
\hline
\end{tabular}
}
\end{table}


\begin{figure}[!p]
\begin{center}
    \includegraphics[width=1\textwidth]{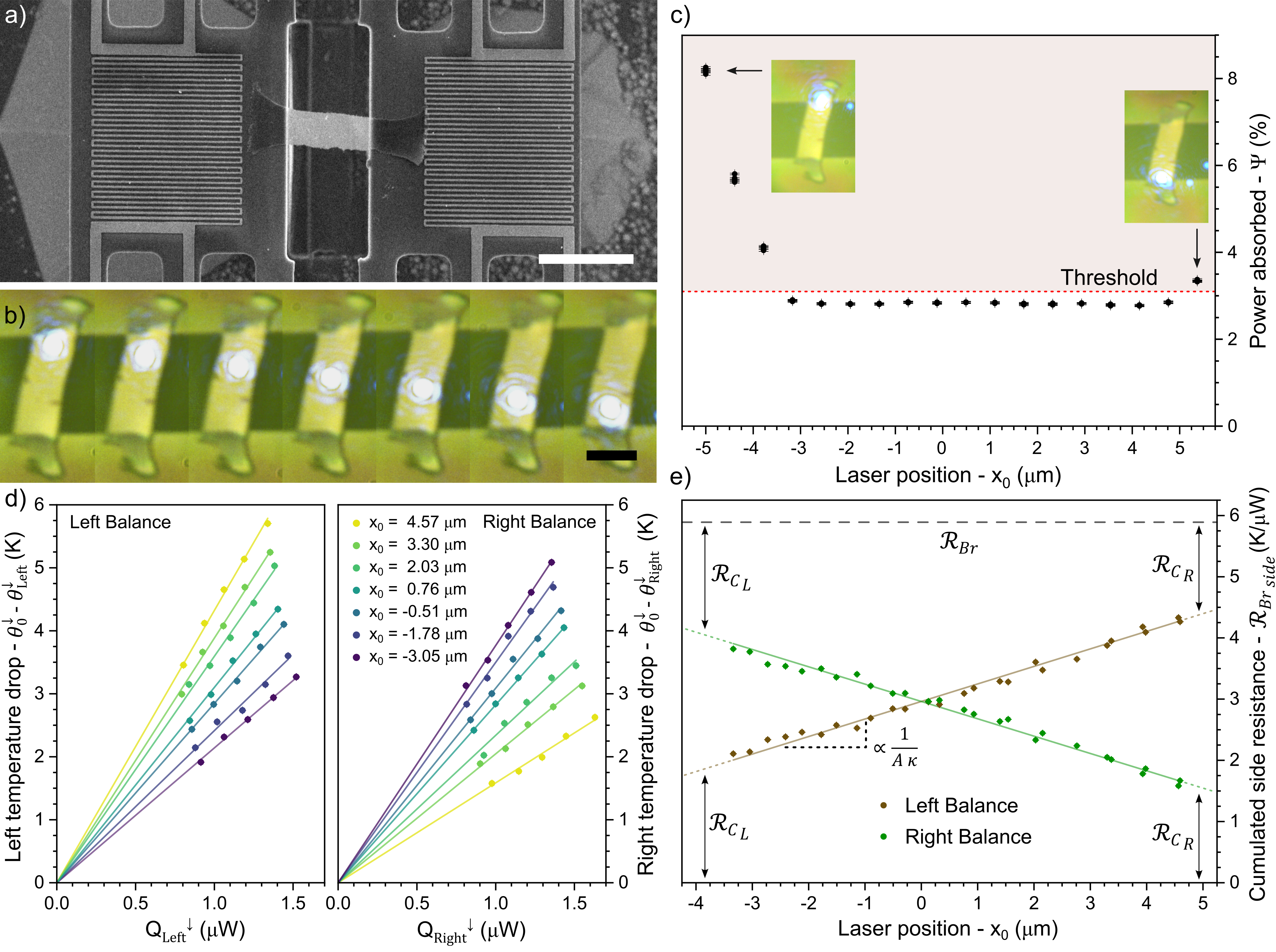}
      \caption{a)~SEM image of a suspended platform device bridged by a membrane. \textcolor{markCol1}{Scale bar is \qty{10}{\micro m}.} 
      b)~Sequence of optical images with the laser beam focused at different positions along the longitudinal axis of the suspended membrane. \textcolor{markCol1}{Scale bar is \qty{5}{\micro m}.} 
      \textcolor{markCol1}{c)~Power absorption measured along the tested membrane. The red dashed line indicates the filter threshold for that deposit non-negligible heat at the nitride membrane.}
      d)~Representative data set for the measured side temperature drop as a function of heat flux, controlled by changing the laser power, at different laser beam position. The experimental data were fitted with the function $(\theta_0^\downarrow-\theta_{side}^\downarrow)/Q_{side}^\downarrow(x)$ at each laser position.
      e)~Representative resulting ${\mathcal{R}_{Br}}_{side} (x_0)$ as a function of laser positions. The fitted slopes of these curves yield the thermal conductivity of the sample. Noteworthy, $x_0$ is arbitrarily offset to 0 at the point where ${\mathcal{R}_{Br}}_{L} = {\mathcal{R}_{Br}}_{R}$.}
      \label{fig:3ProbeResults}
\end{center}
\end{figure} 

The thermal experiment is schematically illustrated in \autoref{fig:3ProbeResults}. Each of the membranes was transferred onto a microdevice featuring a pair of suspended platforms, as depicted in \autoref{fig:3ProbeResults}a. Here, the bridge method was applied prior to the use of the laser source \cite{Swinkels2015,Arya2025}. Subsequently, the thermal balance described in \autoref{sec:3ProbeMethod} was performed at each laser position, as illustrated in \autoref{fig:3ProbeResults}b. The cumulated thermal resistance of the nanostructure defined at each side of the laser position $\mathcal{R}_{\mathrm{Br,side}}(x_0)$ (see \autoref{fig:MethodSketch}b) can be defined as: 
\begin{equation}
\mathcal{R}_{\mathrm{Br,side}}(x_0)
= 
\left. 
\frac{\partial (\theta_0^{\downarrow} - \theta_{\mathrm{side}}^{\downarrow})}
     {\partial Q_{\mathrm{side}}^{\downarrow}}
\right|_{x_0}
= \mathcal{R}_{\mathrm{NS}}\frac{x_0}{L} + \mathcal{R}_{C,\mathrm{side}}, 
\end{equation}

where $\mathcal{R}_{\mathrm{NS}}$ and $\mathcal{R}_{C,\mathrm{side}}$ are the thermal resistance corresponding to the suspended part of the nanostructure and contacts, respectively. \textcolor{markCol1}{ Notably, using $Q_{\mathrm{side}}^{\downarrow} = G_{b,\mathrm{side}}^{\times}\,\theta_{\mathrm{side}}^{\downarrow}$ automatically corrects for variations in laser absorption due to sporadic variations of $P_0$ or sample inhomogeneities along the longitudinal direction. Furthermore, the absorption of the laser beam, measured at each point as $\psi (x_0) = \left(Q_L^{\downarrow} + Q_R^{\downarrow}\right)/{P_0}$, is used to filter out points where part of the beam heat is deposited directly on the nitride platforms. An absorption threshold slightly above the average measured in the center of the membrane is used for this purpose (see \autoref{fig:3ProbeResults}c).} In order to improve the accuracy of the measurement, several laser powers were used at each laser position. Then, the variation of the measured temperatures as a function of the heat flows, which are directly proportional to the laser power $P_0$, was fitted with a linear function, as illustrated in \autoref{fig:3ProbeResults}\textcolor{markCol1}{d}. 
Finally, differentiation with respect to the laser position yielded a relation that is independent of $\mathcal{R}_{C,\mathrm{side}}$:
\begin{equation}
\frac{\partial}{\partial x_0}
\left(
\frac{\partial (\theta_0^{\downarrow} - \theta_{\mathrm{side}}^{\downarrow})}
     {\partial Q_{\mathrm{side}}^{\downarrow}}
\right)
=
\frac{\partial \mathcal{R}_{\mathrm{Br,side}}(x_0)}{\partial x_0}
= \frac{\mathcal{R}_{\mathrm{NS}}}{L}
= \frac{1}{A\kappa},
\end{equation}

where the left-hand side was obtained from the balance data as the slope of the linear fit of the $\mathcal{R}_{\mathrm{Br,side}}(x_0)$ curves represented in \autoref{fig:3ProbeResults}\textcolor{markCol1}{e}. Additionally, contact resistances were extracted by extrapolating $\mathcal{R}_{\mathrm{Br,side}}(x_0)$ to the position $x_c$ where the contact is defined to start, i.e. where conduction is no longer restricted to the membrane cross section used. Thus, at a given $x_c$:
\begin{equation}
\begin{aligned}
\mathcal{R}_{C,R} = \mathcal{R}_{\mathrm{Br},R}(x_{c,R})
&= \mathcal{R}_{\mathrm{Br}} - \mathcal{R}_{\mathrm{Br},L}(x_{c,R}), \\
\mathcal{R}_{C,L} = \mathcal{R}_{\mathrm{Br},L}(x_{c,L})
&= \mathcal{R}_{\mathrm{Br}} - \mathcal{R}_{\mathrm{Br},R}(x_{c,L}).
\end{aligned}
\end{equation}

Remarkably, contrary to the bridge method, with the proposed three-probe approach the estimation of $\kappa$ is not affected by the arbitrary selection of the contact position $x_c$, since $\kappa$ is extracted from the longitudinal gradient of $\mathcal{R}_{\mathrm{Br,side}}(x_0)$. 
In the present study, $x_c$ were selected where the membranes meet the nitride platforms, i.e. the sample lengths (see \autoref{tab:StrucuturalResults}) were defined by the gap between the platforms so that it matches the definition of the thermal resistance obtained from the bridge method $\mathcal{R}_{\mathrm{Br}}$. \textcolor{markCol1}{Interestingly, none of the samples exhibited significant deviations from the linear $\mathcal{R}_{\mathrm{Br,side}}(x_0)$ relationship, as exemplified in \autoref{fig:3ProbeResults}e. Since SEM analysis (\autoref{fig:3ProbeResults}a) revealed no significant variations in sample width, this indicates that $\kappa$ was homogeneous across the samples at the micron-scale resolution probed by the focused laser spot.}

\autoref{tab:ThermalResults} summarizes the measured thermal resistances for all investigated samples. It is worth noticing that there is a large variability in the contact resistance across different samples, despite they all show comparable contact areas. This highlights the significance of accounting for thermal contact resistance in order to accurately characterize the thermal properties of 2D systems.

\begin{table}[!tb]
\centering
\caption{Summary of the thermal characterization of each sample studied at \qty{300}{K}.}
\label{tab:ThermalResults}
\begin{tabular}{ccccccc}
 & \multicolumn{3}{c}{Thermal resistances} & \multicolumn{2}{c}{Thermal conductivity} \\
Sample & Total & Contact & Membrane & Effective & Intrinsic \\
 & $\mathcal{R}_{Br}$ (\unit{K/\micro W}) & $\mathcal{R}_{C}$ (\unit{K/\micro W}) & $\mathcal{R}_{NS}$ (\unit{K/\micro W}) & $\kappa_{eff}$ (\unit{W/m.K}) & $\kappa_{intr}$ (\unit{W/m.K}) \\
\hline
S0        & \num{3.96(.02)} & \num{2.73(.03)} & \num{1.23(.01)} & \num{46.45(2.75)} & \num{46.45(2.75)} \\
\hline
S1        & \num{5.89(.04)} & \num{3.17(.03)} & \num{2.72(.02)} & \num{19.94(.82)}  & \num{22.34(1.61)} \\
S2        & \num{5.75(.20)} & \num{2.88(.12)} & \num{2.87(.05)} & \num{20.55(1.58)} & \num{23.30(2.43)} \\
S3        & \num{6.97(.09)} & \num{2.70(.04)} & \num{4.27(.03)} & \num{12.68(.81)}  & \num{15.33(1.94)} \\
S4        & \num{16.99(1.28)} & \num{4.48(.35)} & \num{12.5(.28)} & \num{7.29(.85)}   & \num{9.46(1.27)} \\
\hline
\end{tabular}
\end{table}

\begin{figure}[!tb]
\begin{center}
    \includegraphics[width=1\textwidth]{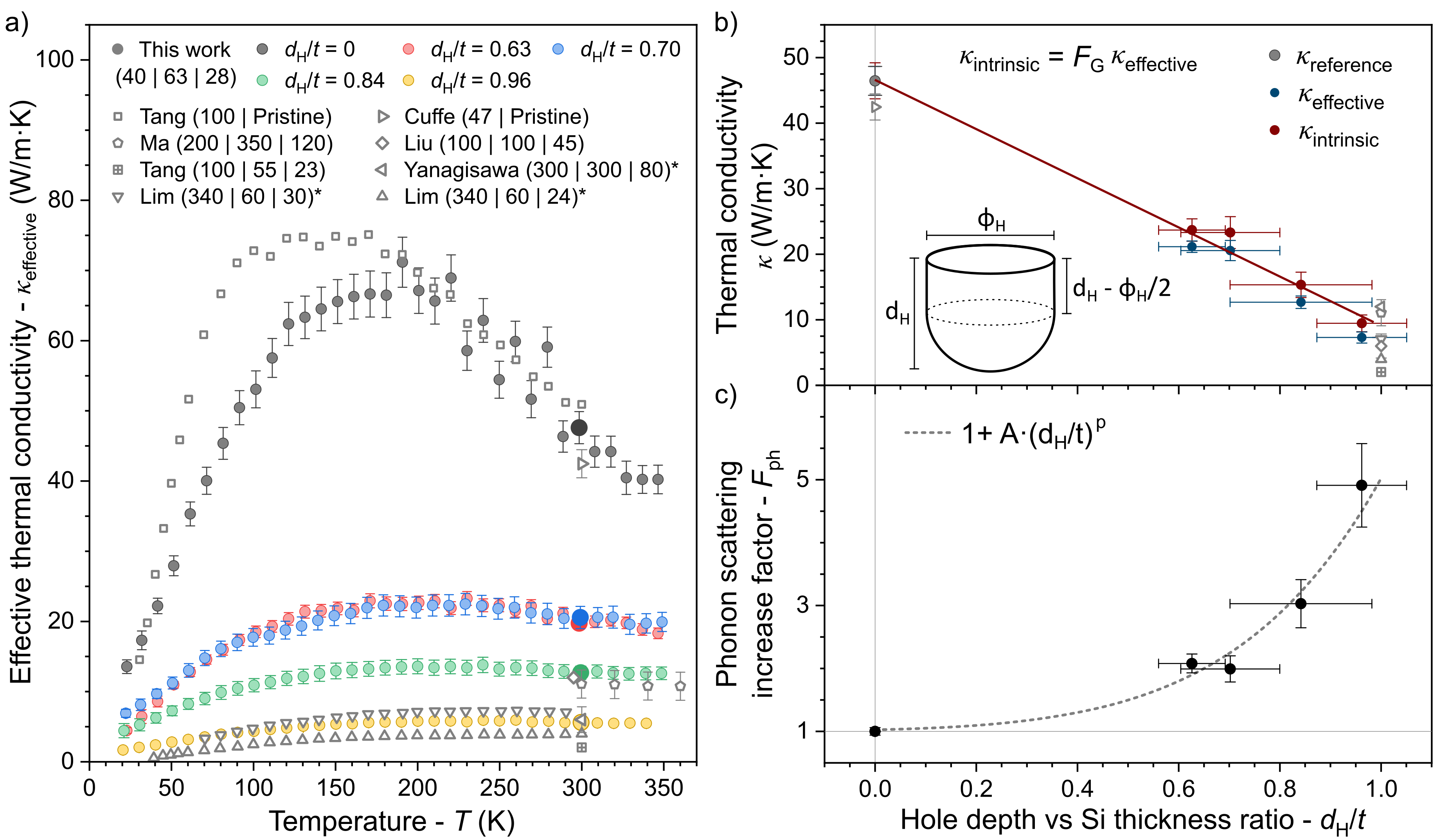}
      \caption{a)~Effective thermal conductivity values using the combined bridge and three-probe methods \textcolor{markCol2}{(solid points). Slightly lighter-filled points represent corrected bridge-method data with $\mathcal{R}_C(300K)$ whereas direct measurement from three-probe experiment at \qty{300}{K} are represented with darker fill color and bigger size}. The thermal conductivity of samples is clearly reduced as the hole depth ratio increases until $d_H/t=0.96$ (practically fully etched through). Literature data is represented in gray \textcolor{markCol2}{empty symbols} \cite{Tang2010,Cuffe2015,Ma2015,Liu2020,Yanagisawa2020,Lim2016}. The legend numbering represents the sizes (thickness~$\vert$~pitch~$\vert$~neck) of the holes in nm for each study. b)~Effective and intrinsic thermal conductivity values at \qty{300}{K} as a function of $d_H/t$. \textcolor{markCol1}{The inset illustrates the geometric modeling of a hole used to calculate the geometric factor $F_G$.} c)~Phonon scattering increase ratio ($F_{ph} = \kappa_{intr} / \kappa_{Ref}$) as a function of $d_H/t$.  
      The terms “effective” and "intrinsic" indicate whether the volume considered for the evaluation of thermal conductivity includes those of the holes or not, respectively.}
      \label{fig:Kappa}
\end{center}
\end{figure}

Finally, the effective thermal conductivities of the films studied with different etching depths are shown in \autoref{fig:Kappa}a. The term effective refers to a definition in which the total volume is considered, including that of the holes\textcolor{markCol1}{, hence $\kappa_{eff} = L/(t W \mathcal{R}_{NS})$}. When comparing these corrected values with those directly obtained from the bridge method, which include the thermal contact resistance, correction factors ($\mathcal{R}_{Br}/\mathcal{R}_{NS}$) of up to 3 in some samples were observed (see \autoref{tab:ThermalResults}), highlighting the significant--sometimes dominant--contribution of the contact resistances in thin film measurements. 
\textcolor{markCol1}{Importantly, corrected thermal conductivities departing from \qty{300}{K} have to be taken with care. Down to $\sim$\qty{180}{K}, $\mathcal{R}_C$ are expected to remain relatively constant as most of phonon modes are well-populated. However, below this temperature, phonon dispersion mismatch can play a significant role, and is expected to increase contact resistances significantly\cite{Monachon2016}. Therefore, at low temperatures (below $\sim$\qty{180}{K}), corrected thermal conductivities are expected to be underestimated.} \textcolor{markCol2}{Indeed, a sensitivity analysis on the effect of the temperature dependency of $\mathcal{R}_C$ (\autoref{fig:KappaSens}) indicates that the differences between reference sample S0 and data reported by \citet{Tang2010} at low temperatures could be explained by this phenomenon. This effect is progressively less relevant in the rest of samples, where the $\mathcal{R}_{NS}$ becomes more dominant over $\mathcal{R}_C$.}

For the pristine sample (S0 with $d_H/t = 0$) a clear temperature dependence was observed. Thermal conductivity initially increased--driven by the specific heat ($\propto T^1$ in this case)\cite{Tang2010}--until it eventually saturated at \qtyrange{150}{170}{K} and decreased again for higher temperatures. This is caused by Umklapp phonon scattering becoming dominant at temperatures above \qty{200}{K}. \textcolor{markCol2}{The results at room temperature match those of \qtyrange{40}{50}{W/m.K} for Si thin films with similar thickness reported by \citet{Tang2010} and \citet{Cuffe2015}, (the open squares and right-pointing triangles in \autoref{fig:Kappa}a respectively), supporting the validation of the three-probe correction for the contact resistance contribution.} \textcolor{markCol1}{However, it is also evident how the corrected values seem to underestimate $\kappa_{eff}$ at the low temperature range ($<$\qty{180}{K}), likely by an increase in $\mathcal{R}_C$--here considered constant--with respect to the values measured at room temperature.}  

All these data illustrate how the phonon confinement in silicon films as thin as \qty{40}{nm} is still insufficient for an effective suppression of phonon transport, reaching values as high as \qty{70}{W/m.K} at intermediate temperatures, in contrast with silicon nanowires of comparable diameters\cite{Li2003, Sojo2022, Lee2016, Sojo2023c, Sojo2024a}. This high propagation is likely driven by \textcolor{markCol1}{the} large mean free path oscillation modes (\qtyrange{50}{5000}{nm}) contained in the unconfined plane, which can travel practically unimpeded\cite{Esfarjani2011,Ziman2001}. 

Therefore, the progressive etch of holes at one of the surfaces is expected to have a large impact on the phonon transport, hindering the propagation of in-plane oscillations. This is clearly confirmed in \autoref{fig:Kappa}a, since a clear trend is observed in the overall magnitude of the thermal conductivity as a function of the defined $d_H/t$ ratio, which decreases with increasing ratios. The additional phonon scattering in the in-plane propagation eventually becomes the dominant mechanism at all frequencies, drastically suppressing the thermal conductivity and reaching a value of \qty{7.29(.85)}{W/m.K} at room temperature for the practically fully etched holes (S4 with with $d_H/t = 0.96$). Noteworthy, this reduction represents a 6-fold decrease in the \textcolor{markCol1}{effective} thermal conductivity achieved through nanostructuration. This prevalence of the boundary scattering shifts the maximum of the thermal conductivity to higher temperatures, analogously to the effect observed by \citet{Li2003} and \citet{Tang2010} in silicon nanowires and holey thin films, respectively. Remarkably, the thermal conductivity found for sample S4 approaches the value of \qty{2.03}{W/m.K} reported by \citet{Tang2010} in thin films with comparable hole, thickness and pitch sizes. However, in this referenced study, the lack of a characterization of the contact resistance likely underestimated the true value of the sample thermal conductivity. \textcolor{markCol1}{Noteworthy, the low-temperature behavior of this sample is in better agreement with the literature than that of the pristine sample (S0). This is likely because the phonon suppression caused by nanostrucutration shifts the sample's phonon dispersion closer to that of the nitride, minimizing the effect of phonon dispersion mismatch at low temperatures.}

\textcolor{markCol1}{Effective} thermal conductivity may also be compared in terms of film porosity. Based on the STEM characterization (\autoref{fig:STEM})\textcolor{markCol1}{, porosity $\varrho$ and the geometric factor $F_G$--which directly accounts for the effective reduction of the cross section of the silicon--can be estimated by a simple modeling of the hole geometry. Holes are modeled as semi-spheres of diameter $\phi_H$ coupled with a cylindrical part of height $(d_H-\phi_H⁄2)$, as depicted in the inset of \autoref{fig:Kappa}b. Therefore, the individual volume of a hole is: 
\begin{equation}
    V_H = \frac{2}{3}\pi\left(\frac{\phi_H}{2}\right)^3 + \pi\left(d_H - \frac{\phi_H}{2}\right)\left(\frac{\phi_H}{2}\right)^2
    \label{eq:HoleVolume}
\end{equation}
Hence, the porosity and the geometrical factor are simply: 
\begin{equation}
    F_G = \frac{1}{1 - \varrho} = \frac{t \cdot A}{v_{Si} \cdot A} = \frac{t}{v_{Si}} = \frac{t}{t - v_H}
    \label{eq:GeometricFactor}
\end{equation}
where $v_{Si}$ and $v_H=\rho_H·V_H$ are the solid Si and hole volumes per unit area, respectively. Notice that $v_H$ uses the measured hole density $\rho_H$ per unit area and the estimation of the hole volume (\autoref{eq:HoleVolume}). The film thickness can therefore be used directly, as it represents a full volume of thinness $t$ per unit area.} 

\textcolor{markCol1}{Uncertainty in the calculation of both $F_G$ and $\varrho$ is propagated from the STEM characterization (\autoref{fig:STEM}) assuming each measure quantity to be independent. The holes patterned in the films} have a pitch of $\sim$\qty{63}{nm} and an average diameter of \qty{35(3)}{nm} (the BCP pattern originally has a diameter of $\sim$\qty{25}{nm}--see \autoref{sec:Expermental}--but the pattern transfer process enlarges it by \qty{10(3)}{nm}). \textcolor{markCol1}{Hence, a maximum $F_G$ \num{1.30(0.09)} was calculated for the practically fully etched holey film (S4, with $d_H/t =$ \num{0.96(9)}). Equivalently,} the porosity for these films is estimated to be \textcolor{markCol1}{\num{22.9(0.4)}\%, as opposed to the films reported by \citet{Tang2010} which had a value of 35\%.} This difference in porosity could partially explain the differences in \textcolor{markCol1}{$\kappa_{eff}$} as this effect has also been observed in porous silicon nanowires with comparable porosity\cite{Yang2021}. On the other hand, \citet{Lim2016} reported values between 4 and \qty{7}{W/m.K} for holey Si films with hexagonal patterns and neck sizes between 24 and \qty{30}{nm}, which are in good agreement with those obtained in this work (average neck sizes of $\sim$\qty{27}{nm}). \citet{Liu2020} also reported a value of \qty{6.00(1.83)}{W/m.K} on a square-like holey film with a neck size of \qty{23}{nm}. Other holey films (namely \citet{Yanagisawa2020, Ma2020}) show higher \textcolor{markCol1}{$\kappa_{eff}$}, likely due to their larger neck sizes (120 and \qty{60}{nm}, respectively). Furthermore, Monte Carlo simulations by \citet{Hao2009} reported that a \qtyrange{45}{50}{nm} thick holey film with a square-like pattern and a porosity of 25\% should have a $\kappa_{Holey}$/$\kappa_{Bare}$ ratio $\sim$\qtyrange{0.11}{0.12}{}. The $\kappa_{Holey}$/$\kappa_{Bare}$ ratio of the films obtained in this work is 0.15, indicating good agreement despite the hole pattern is not a square array but a hexagonal one. In addition, the simulations by \citet{Xie2018} show values around \qtyrange{5}{6}{W/m.K} on a holey film with a period of \qty{60}{nm} and neck sizes between 25 and \qty{27.5}{nm}.

Regarding the intermediate samples (i.e., S1 to S3), the dependence of \textcolor{markCol1}{$\kappa_{eff}$} on the $d_H/t$ ratio shown in \autoref{fig:Kappa}b highlights \textcolor{markCol2}{a good} tunability of \textcolor{markCol1}{$\kappa_{eff}$} through adjustments of $d_H/t$ in the films. The comparatively smaller difference observed between S1 and S2 can be attributed to variations in film thickness, which compensates for the deeper etch fraction in S2 (see \autoref{tab:StrucuturalResults}). \textcolor{markCol1}{In general, a clear control of the thermal conductivity was demonstrated via the etching ratio, as illustrated in \autoref{fig:Kappa}b.} Indeed, a good linear relationship is found, as illustrated by the \textcolor{markCol1}{linear} fit.

\begin{figure}[!tb]
\begin{center}
    \includegraphics[width=0.55\textwidth]{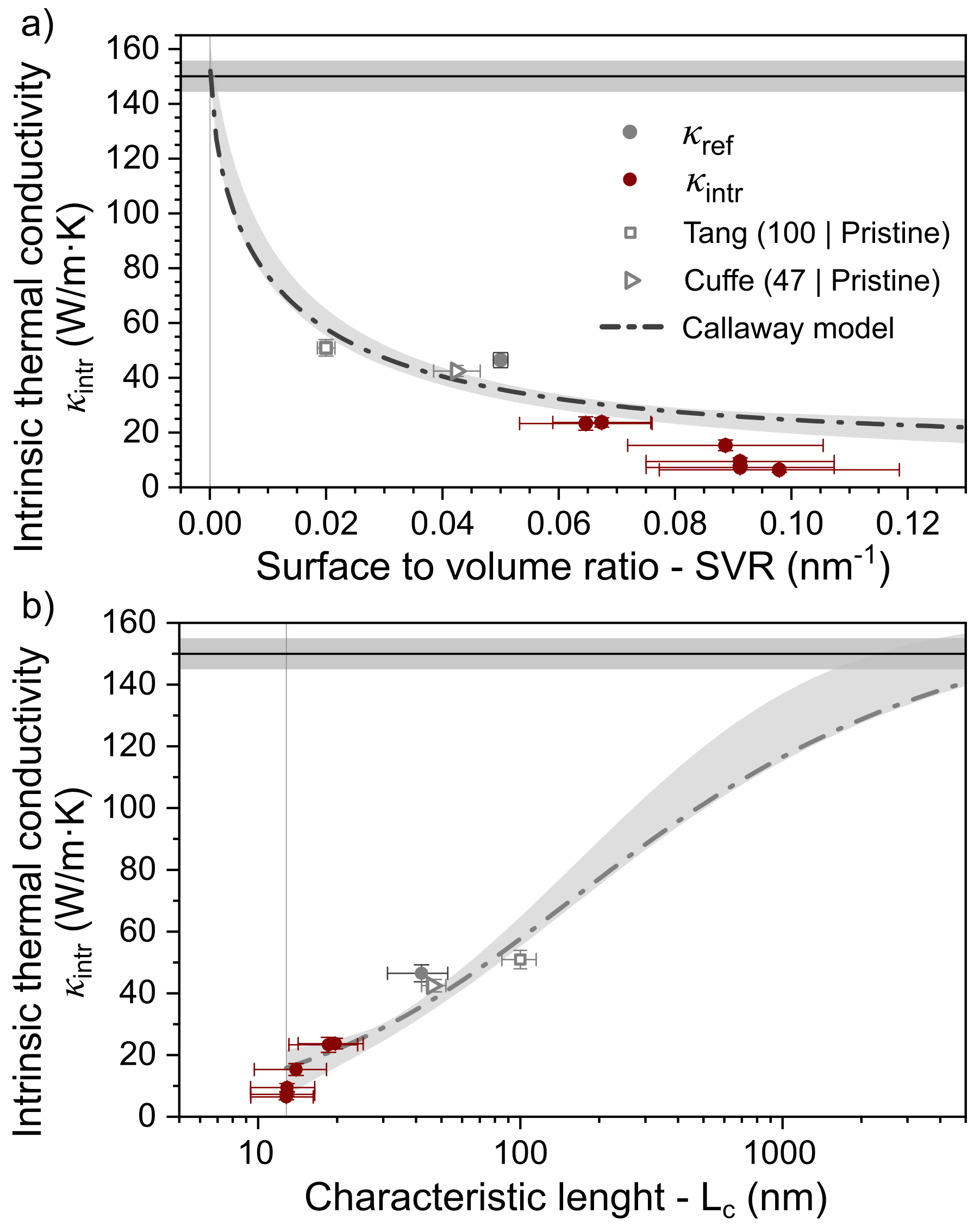}
      \caption{\textcolor{markCol1}{a)~Intrinsic thermal conductivity values at \qty{300}{K} as a function of the surface-to-volume ratio (SVR). The vertical line marks $SVR = 0$ b)~Intrinsic thermal conductivity values as a function of a characteristic length. The vertical line references the neck size ($L_c = n$). In both panels the horizontal solid line represents bulk value.}}
      \label{fig:KappaModel}
\end{center}
\end{figure} 
Another important aspect to elucidate is the effect of nanostructuring on phonon scattering. There are two contributions involved in the reduction of the thermal conductivity: the geometric factor $F_G$ arising from the reduction of the solid fraction on the meta-material volume, and the phonon scattering factor $F_{ph}$. \textcolor{markCol1}{ With the geometrical factor, the definition of the intrinsic thermal conductivities is direct: $\kappa_{intr} = F_G \kappa_{eff}$. For the fully etched holey Si film--i.e. the one defined only to the solid volume--this yields an intrinsic thermal conductivity of $\sim$\qty{9.4}{W/m.K}.}

\textcolor{markCol1}{After excluding the effect of $F_G$, the phonon scattering factor can be calculated as $F_{ph} = \kappa_{intr} / \kappa_{Ref}$ and studied as a function of the $d_H/t$ ratio} (see \autoref{fig:Kappa}c). It becomes clear that increasing $d_H/t$, increases phonon scattering, reaching a maximum of $F_{ph}$ of $\sim$ 5 for $d_H/t = 1$. \textcolor{markCol1}{Empirically, a correlation for $F_{ph}$ could be captured using a power law fit, $F_{ph} = A \cdot (d_H/t)^p$, where $A$ and $p$ are only fitting parameters (\autoref{fig:Kappa}c). We find a power coefficient of $\sim 4$ using the $d_H/t$ ratio, as compared with a roughly $p \sim 1$ trend found in nanowire systems using diameter as the characteristic parameter\cite{Li2003, Mingo2003} or a $p \sim 2$ found in holey silicon when varying the neck size of the pattern\cite{Lim2016}. This indicates a very strong scattering effect that possibly goes beyond the Casimir limit. The irregular partially etched top side may act as a very effective virtual rough surface that traps phonons that deviate from the in-plane continuous layer underneath. Indeed, we observed a strong correlation of $\kappa_{intr}$ with the surface-to-volume ratio (SVR), as observed in \autoref{fig:KappaModel}a. A simple model based on the Callaway/Klemens framework described by \citet{Ohishi2015} integrating an additional semi-empiric SVR scattering rate suited for such high effective roughness \cite{Sojo2023a} was implemented (the dashed line in \autoref{fig:KappaModel}a). A simple numerical stochastic variation of the input parameters with uncertainty generates the 95\% confidence interval. Interestingly, the model captures the overall phenomena, even when using an empirically calculated parameter for such scattering term extracted from rough Si NW data \cite{Yang2020}. However, it is clear that phonon scattering at high SVR is underestimated using this simple modeling. More refined approaches such as Monte Carlo phonon tracing \cite{Anufriev2024} or finite element simulations implementing a kinetic collective model \cite{Beardo2019} would be required to capture the observed phenomena more accurately.}

\textcolor{markCol1}{A final analysis (\autoref{fig:KappaModel}b) is conducted analyzing the data as a function of an effective characteristic length defined as $L_c = \sqrt{\max(t - d_H, 0)^2 + n^2}$ with $n$ the neck size measured in \autoref{fig:STEM}c. In pristine thin films, the $L_c$ is simply the thickness of the film $t$. Then, as the holes are progressively etched, we considered the thickness of the channel underneath the holes ($t - d_H$). However, as the hole depth starts reaching the opposite side of the film, this channel narrows, until at some point the average neck size between holes remains as the largest feature of the structure. A clear trend is also observed, with a decreasing $\kappa_{intr}$ with smaller $L_c$. Here, a saturation appears for samples S3 and S4 as their etching is almost completed and their $L_c$ tends to the neck size. The same simple model based on the Callaway/Klemens framework is applied here too (dashed line in \autoref{fig:KappaModel}b). Again, while the overall thermal conductivity decrease is reasonably captured, further sophisticated modeling for the low $L_c$ cases would be required.}

\section{Conclusions}

Nanostructured films with \qty{63}{nm} pitch cylindrical patterns were successfully fabricated using fully scalable self-assembly process of PS-b-PMMA BCPs through blending. Subsequently, these patterns were transferred onto Si films using both wet and dry etching techniques, demonstrating that PS-b-PMMA nanometric patterns can be directly transferred (without any intermediate layer) onto \qty{40}{nm} thick Si films with \textcolor{markCol1}{an adequate} anisotropy and reproducibility. \textcolor{markCol2}{The depth of the hole pattern can be reliably tuned by adjusting the reactive ion etching time.}. After etching, holes of $\sim$\qty{35}{nm} in diameter were obtained. \textcolor{markCol1}{The integration of these membranes onto suspended microplatforms for thermal bridging measurements using micro manipulation was also demonstrated.} 

Moreover, based on finite element analysis, the thermal conductivity measurements on 2D membranes using a three-probe method with a laser source were confirmed to be accurate for samples with an aspect ratio \textcolor{markCol1}{$W/L\le0.5$} and a suspended length 10 times larger than the beam size. The method was validated on pristine \qty{40}{nm} thick reference silicon films, obtaining values (\qty{46.45(2.75)}{W/m.K}) matching both experimental and theoretical estimations at room temperature. Nevertheless, the variability in thermal contact resistance observed across the different samples highlights the critical need to account for this effect in order to achieve a reliable thermal evaluation. In addition, it is worth noting that for temperature dependent conductivities, the confidence of the results decreases as the temperatures depart from \qty{300}{K}, since a constant resistance value was used due to experimental limitations. Further exploration of the behavior of the thermal contact resistance as a function of temperature will result in even more accurate measurements. 

For films patterned with the described cylindrical geometry, effective thermal conductivity measurements demonstrated a 6-fold decrease. The measured thermal conductivity of the practically etched-through film was \qty{7.29(.85)}{W/m.K} at room temperature, comparable to similar reported films fabricated with more complex approaches (e-beam lithography or block copolymer lithography with intermediate masks). \textcolor{markCol2}{Remarkably, a clear correlated linear decrease in thermal conductivity was observed with increasing etching depth. This degree of control is particularly appealing given the intrinsic scalability of the process.} Moreover, these patterned films can be integrated into a CMOS- and wafer-compatible device fabrication process to selectively nanostructure silicon films in specific regions by simply employing a blocking hard mask. Hence, an interesting pathway for the realization of advanced thermal dissipation is opened. This could be exploited, for instance, in thermoelectrics, where the 5-fold reduction in the \textcolor{markCol1}{intrinsic} thermal conductivity with a minor $\sim$1.3 reduction of the structural solid volume should have a minor impact on the electronic properties, thus boosting the $zT$ factor. 

In summary, by controlling the depth of the patterned holes, \textcolor{markCol2}{an evident modulation of the effective thermal conductivity is revealed}, demonstrating geometry-driven thermal transport engineering in silicon thin films. These results establish a scalable route for phonon engineering using block copolymer patterning, \textcolor{markCol2}{while also showcasing the optical three-probe technique as a powerful tool for quantitative, spatially resolved thermal conductivity measurements in complex thin-film systems.}

\appendix
\setcounter{figure}{0}
\renewcommand{\thefigure}{A\arabic{figure}}
\renewcommand{\thetable}{A\arabic{table}}

\section*{ \textcolor{markCol2}{ Appendix: Sensitivity analysis of the $\mathcal{R}_C$ correction} }

\textcolor{markCol2}{
To further understand the validity of the approximation of using a constant $\mathcal{R}_C$, a sensitivity analysis is performed. The relative absolute change in the corrected effective thermal conductivity of the most extreme cases (S0 with $d_H/t = 0$ and S4 with $d_H/t = 0.96$) were studied after a 5\% change in the $\mathcal{R}_C$. Two scenarios are considered: (i) A $\mathcal{R}_C$ that is temperature independent ($\propto T^0$) and; (ii) a $\mathcal{R}_C$ that empirically follows a $\mathcal{R}_1 + \mathcal{R}_2 /T$ trend that  grows at low temperatures ($\propto T^{-1}$) a as expected from the known theory \cite{Monachon2016}. The values of of the high temperature constant $\mathcal{R}_1$ and low temperature factor $\mathcal{R}_2$ were set to match the experimental $\mathcal{R}_C$ of each sample at \qty{300}{K} (see \autoref{tab:ThermalResults}). 
}

\begin{figure}[!tb]
\begin{center}
    \includegraphics[width=0.55\textwidth]{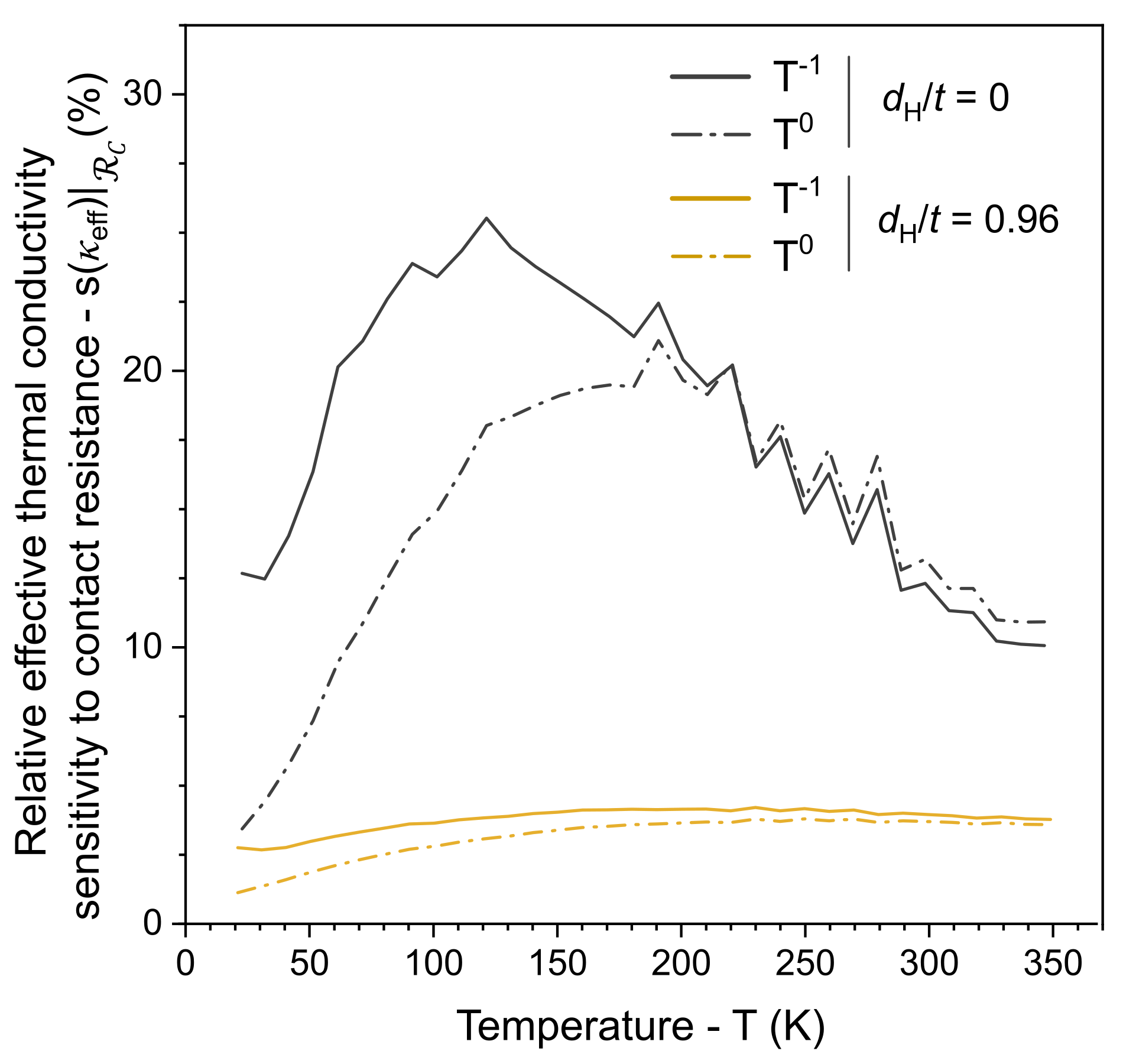}
      \caption{\textcolor{markCol2}{Relative sensitivity of the effective thermal conductivity $\kappa_{eff}$ for a 5\% variation in the thermal contact resistance $\mathcal{R}_C$, assuming either a temperature-independent value ($\propto T^0$) or a temperature dependent empirical trend $\mathcal{R}_1 + \mathcal{R}_2 /T$ ($\propto T^{-1}$), for samples S0 ($d_H/t = 0$) and S4 ($d_H/t = 0.96$).}}
      \label{fig:KappaSens}
\end{center}
\end{figure} 

\textcolor{markCol2}{
As it can be appreciated in \autoref{fig:KappaSens}, in the first scenario with temperature independent $\mathcal{R}_C$, the effect on the low conductive sample (S4 with $d_H/t = 0.96$) is minimal, because $\mathcal{R}_{NS}$ is the limiting one. However, the effect is significantly larger in the reference sample (S0 with $d_H/t = 0$), where the contacts account for a larger proportion of the total resistance (see \autoref{tab:ThermalResults}). In such cases, the maximum deviation appears at the peak of $\kappa_{eff}$ (where  $\mathcal{R}_{NS}$ is minimal), and the error can propagate as high as 20\%. On the second scenario, where the $\mathcal{R}_C$ is set to grow with a $T^{-1}$ trend at low temperatures, one can observe the magnitude of the deviation has maximum at \qtyrange{70}{140}{K}, and the $\kappa_{eff}$ would be higher in this range. Interestingly, this fairly agrees with the discrepancies found with the data of \citet{Tang2010} (\autoref{fig:Kappa}a), supporting the notion that assuming a temperature-independent $\mathcal{R}_C$ leads to the largest deviation in this temperature range. However, this analysis should only be interpreted in qualitative terms, as the resulting errors are highly dependent on the actual trend of $\mathcal{R}_C$ with temperature.
}

\begin{acknowledgement}
This work has been supported by the Research Fund of the University of Basel for Excellent Junior Researchers (project \textit{GATEWAY}). This project has also received funding from the Swiss National Science Foundation (Grant No. \textbf{189924}) project \textit{HYDRONICS} and from the European Research Council (ERC) under the European Union's Horizon 2020 research and innovation program (Grant Agreement \textbf{756365}). This work was also supported by the project \textit{nanoDecoTEG} (grant \textbf{TED2021-129612B-C21}) and grant \textbf{RYC2022-037016-I}, funded by the Spanish Agency of Research (AEI) program (\textbf{10.13039/501100011033}) by the European Union NextGeneration EU/PRTR and ESF+. The authors also acknowledge partial funding from the María de Maeztu programme (\textbf{CEX2023-001397-M}), also financed by the Spanish Agency of Research (AEI) program (\textbf{10.13039/501100011033}). Financial support from the Agència de Gestió d’Ajuts Universitaris i de Recerca (AGAUR) of the Generalitat de Catalunya under grant \textbf{2021 SGR 00497} is gratefully acknowledged as well. This research has made use of the Spanish ICTS Network \textit{MICRO-NANOFABS} clean room facilities, partially funded by FEDER funds through the \textit{MINATEC-PLUS-2} project (\textbf{FICTS2019-02-40}).

\end{acknowledgement}

\section*{Author contributions}
\textbf{J.M.S.G.:} conceptualization, methodology, supervision, investigation, data curation, formal analysis, visualization, software, writing - original draft;
\textbf{A.R.I.:} conceptualization, investigation, data curation, formal analysis, visualization, writing - original draft;
\textbf{D.M.K.:} verification, investigation, writing - review;
\textbf{A.N.} verification, investigation;
\textbf{I.M.F.} conceptualization, writing - review;
\textbf{M.F.R.} conceptualization, funding acquisition, writing - review;
\textbf{M.S.} conceptualization, funding acquisition, writing - review;
\textbf{I.Z.} conceptualization, funding acquisition, writing - review.




 \small{\bibliography{References}}



\end{document}